# *In situ* kinetics studies of Zn-Al LDH intercalation with corrosion related species

Mariia H. Iuzviuk[a,*], Anissa C. Bouali[b], Maria Serdechnova[b], Kiryl A. Yasakau[c], D.C. Florian Wieland[b], Gleb Dovzhenko[b], Aliaksandr Mikhailau[c], Carsten Blawert[b], Igor A. Zobkalo[a], Mario G.S. Ferreira[c], Mikhail L. Zheludkevich[b, d]

[a] Petersburg Nuclear Physics Institute, Laboratory of Physics of Crystals 188300, Leningradskaya Oblast, Gatchina, 1, mkr. Orlova Roshcha, Russia.

[b] Institute of Materials Research, Helmholtz-Zentrum Geesthacht, Max-Planck-Straße 1, 21502 Geesthacht, Germany.

[c] CICECO - Aveiro Institute of Materials, Dep. Materials and Ceramic Engineering, University of Aveiro, 3810-193 Aveiro, Portugal.

[d] Faculty of Engineering, Kiel University, Kaiserstraße 2, Kiel 24143, Germany

[*] Corresponding author: Tel./Fax: +7 81371 46416/36025 email: yuzyvuk_mh@pnpi.nrcki.ru

**Abstract**

Kinetics parameters for three anion exchange reactions - Zn-LDH-NO$_3$ → Zn-LDH-Cl, Zn-LDH-NO$_3$ → Zn-LDH-SO$_4$ and Zn-LDH-NO$_3$ → Zn-LDH-VO$_x$ - were obtained by *in situ* synchrotron study. The first and the second ones are two-stage reactions; the first stage is characterized by the two-dimensional diffusion-controlled reaction following deceleratory nucleation and the second stage is a one-dimensional diffusion-controlled reaction also with a decelerator nucleation effect. In the case of exchange NO$_3^-$ → Cl$^-$ host anions are completely released, while in the case of NO$_3^-$ → SO$_4^{2-}$ the reaction ends without complete release of nitrate anions. The exchange of Zn-LDH-NO$_3$ → Zn-LDH-VO$_x$ is one-stage reaction and goes much slower than the previous two cases. It is characterized by a one stage two-dimensional reaction and nucleation considered to be instantaneous in this case. As a result, at the end of this process there are two crystalline phases with different polyvanadate species, presumably V$_4$O$_{12}^{4-}$ and V$_2$O$_7^{4-}$, nitrate anions were not completely released. The rate of replacing NO$_3^-$ anions by guest ones can be represented as Cl$^-$ > SO$_4^{2-}$ > VO$_x^{y-}$.

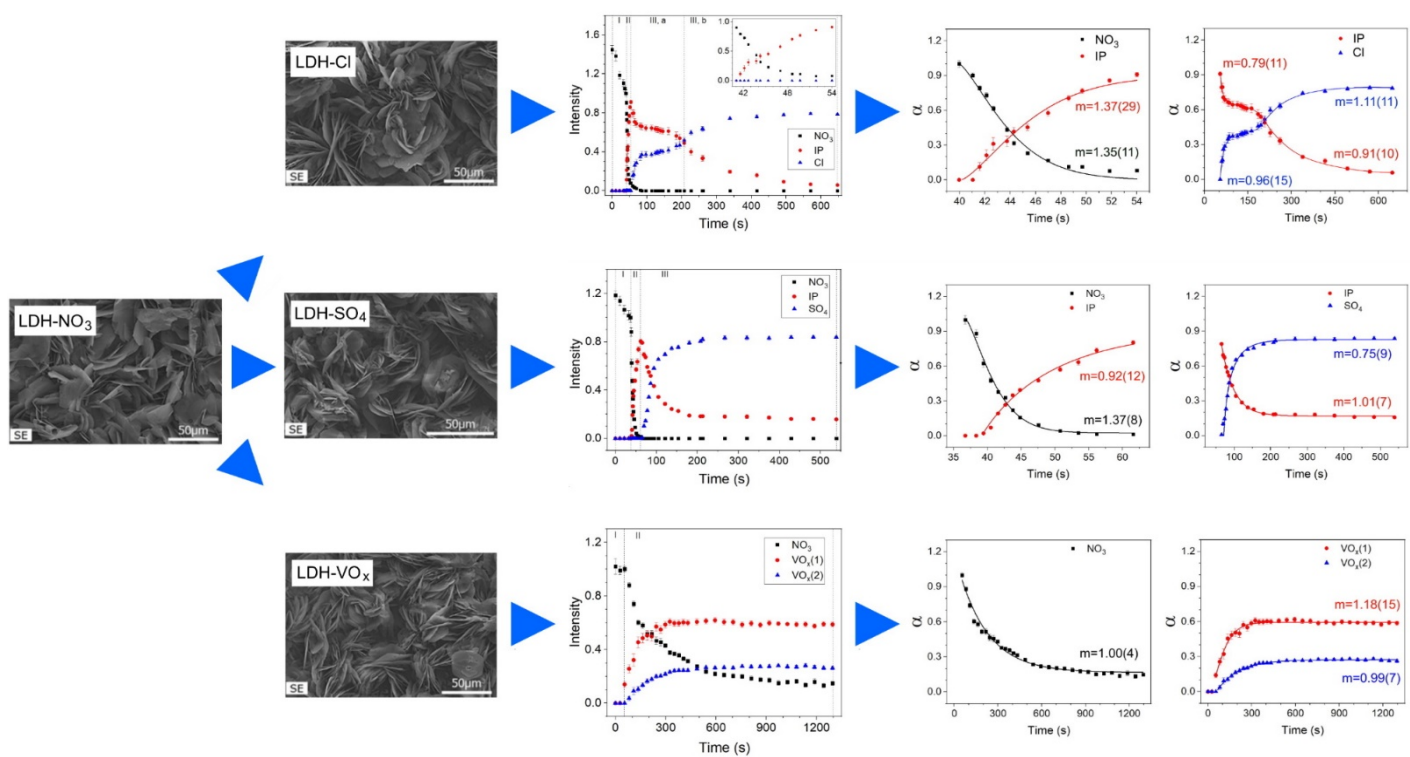

## 1. Introduction

Layered double hydroxides (LDHs), also known as anion exchanging clays, have attracted a considerable interest in the last decades. LDHs are regarded as promising materials for a broad number of practical applications due to their ability to be tuned for specific functionalities such as the intercalation of different types of species (inorganic, organic, biomolecules), unique structure, simplicity of synthesis and others [1–5].

The crystal structure of LDH was first studied by powder X-ray diffraction [6, 7]. Generally it can be represented by the formula $[M^{II}_{1-x}M^{III}_{x}(OH)_2]^{x+}(A^{y-})_{x/y} \cdot zH_2O$, where $M^{II}$ and $M^{III}$ are divalent and trivalent metal cations, respectively, and $A^{y-}$ is an y-valent anion. Cations like $Zn^{2+}$ and $Mg^{2+}$ are mostly used as $M^{II}$ and $Al^{3+}$ is quite popular as $M^{III}$. The mixed hydroxides of these metal cations form layers possessing an overall positive charge, which is balanced by exchangeable anions. The compounds also contain various amounts of water molecules, hydrogen bonded to the hydroxide layers and/or to the interlayer anions. The crystal structure of the layered compounds can have several variations depending on $M^{II}$ and $M^{III}$ cations choice and their ratios [8]. The additional flexibility is given by the reversible ability to intercalate different type of anions $A^{y-}$, such as various inorganic [9, 10], organic [11] or biomolecules [12] species.

These features are the base for the great potential of LDH in the design of high-performance functional materials on the nanometer scale. Since LDH compounds can be produced quite easily by hydrothermal synthesis, these materials now are of interest for many applications, such as catalysis [13], fire retardants [14], pharmaceuticals [15], photochemistry/electrochemistry [16], etc.

In recent years, LDHs became also promising functional materials for self-healing protective coatings. After the pioneering work of Buchheit et al. [17], numerous studies revealed an improvement of the corrosion protection performance by the use of LDHs on Al, Zn and Mg alloys [18–23] as well as on anodized surfaces [24, 25] and on galvanized steels [26]. The anion-exchange ability of LDH offers the possibility to store inhibitor anions in the galleries between the hydroxide layers and to release them "on demand" – e.g. triggered by an increase of pH or the presence of aggressive species in solution [27]. The aggressive species, such as $Cl^-$ and $SO_4^{2-}$, are absorbed and trapped between the LDH galleries because of their strong affinity [28] whereas the corrosion inhibitors are released [26, 29].

However, a number of important details of the structure such as the ordering of the metal cations, the stacking arrangement of the layers and the arrangement of anions and water molecules in the interlayers are still not clearly understood. Thus, these points are the subject of controversy in literature [30].

The aim of this work is to study the kinetics of anion exchange reactions in Zn-Al LDHs, grown directly on a pure zinc substrate. Nitrate $NO_3^-$-containing LDH (Zn-Al LDH-$NO_3$) was chosen as the parent compound for further intercalations. The synthesis of LDH-$NO_3$ with various cations is a relatively straightforward process, since the $NO_3^-$ anion has a lower affinity [29], therefore can be easily intercalated with the desired anion by anion exchange reactions. Chloride (Cl$^-$), sulphate (SO$_4^{2-}$) and vanadate (VO$_x^{y-}$) anions were used as guest species during the anion exchanges.

Since fast data acquisition is needed to obtain information on the kinetics, synchrotron radiation was employed for *in situ* studies of the exchange processes. Due to the high flux available at such X-ray sources a very fast measuring of complete XRD patterns can be performed during heterogeneous reactions. The obtained time dependence of the integrated Bragg intensities allows to determine quantitatively the kinetic parameters of the reactions. During these reactions, different crystalline phases could emerge including intermediate one(s), and this technique permits to observe and refine their structure. Both qualitative and quantitative information regarding the mechanism and kinetics of anion exchange reaction can be obtained.

## 2. Experimental
### 2.2. Methods
#### 2.2.1. LDH growth on zinc and anion exchange conditions.

The synthesis procedure employed in this work was reported in a previous publication [31]. Briefly, Zn-Al LDH-nitrate was grown on zinc substrate (>99%) in 1 mM solution of Al(NO$_3$)$_3$ (>99%, Sigma-Aldrich, Germany) and 0.1 M NaNO$_3$ (>99.95%, Sigma-Aldrich, Germany) at the temperature of 90 °C for 20 h.

The anion exchange reactions were carried out using solutions of 0.1 M NaCl (96%, AlfaAesar, Germany), 0.1 M Na$_2$SO$_4$ (99%, Merck KGaA, Germany) and 0.1 M NaVO$_3$ (96%, AlfaAesar, Germany) for 30 min at room temperature. No pH adjustments were made.

#### 2.2.2. SEM and EDS analysis

The study of parental (Zn-Al LDH-NO$_3$) and the final compositions (Zn-Al LDH-Cl/SO$_4$/VO$_x$) was carried out using a Hitachi SU-70 scanning electron microscope (SEM) supplied with Bruker XFlash 5010 energy dispersive spectrometer (EDS) at 15kV accelerating voltage. EDS results are presented in the form of X-ray energy distribution spectra and two-dimensional element distribution (elemental map).

#### 2.2.3. Synchrotron X-ray diffraction.

The study of the intercalation of LDH films on zinc substrates was carried out at the P08 high-resolution diffraction beamline of the PETRA III synchrotron radiation source (DESY, Hamburg, Germany) with an X-ray energy of 25 KeV [32]. The beam size was 2µm x 30 µm focused by compound refractive lenses, incident angle value was Θ=0.12°. The experimental set-up consists of an *in situ* flow cell with an X-ray transparent window allowing the measurement of grazing incident diffraction as described in detail in [33].

The patterns were collected before the start of the anion exchange reaction and then continuously recorded at an interval of 0.54 seconds after the start of the anion exchange reaction.

A two-dimensional PERKIN Elmer detector with a pixel size of 200 µm with a distance of 1.426 m to the sample was used. The resulting 2θ range was 2–19.9°. The radial integration was performed using GSAS II package. Then FAULTS software [34] was utilized for the refinement of the crystal structure and AMORPH program [35] was used for the estimation of the amount of amorphous phases.

## 3. Results and discussion
### 3.1 SEM and EDS results

SEM images along with elemental maps of a parent LDH-NO$_3$ sample and three final samples with varying compositions (LDH-Cl/SO$_4$/VO$_x$) are shown in Fig. 1. EDS analysis is presented as two-dimensional maps of the element distribution. Integral EDS spectra for each composition are shown in Fig. 2. SEM images display typical patterns of LDH conversion coatings [33]. The LDH flakes are typically about 50-100 nm thick, with mostly hexagonal shape. These flakes cover the whole field of view with a certain degree of preference to perpendicular orientation to the substrate. Some crystals exhibit a rosette type growth morphology with a common nucleation site. Roughly, the lateral size of the flakes varies from 30 to 50 µm, which seem to be bigger comparatively to other LDH conversion coatings of the same kind (ZnAl- or MgAl-based), grown on Mg and Al [26, 33]. It can also be seen from SEM images that the morphology of the surface does not change significantly after the anionic exchange reaction. The flakes of LDH preserve their shape, show no damage, delamination or detaching from the surface. Since the sample surface is highly uneven, a significant part of the EDS signal will be dictated by the topography, thus for a better understanding the elemental map series are accompanied with secondary electron SEM images. The distribution of metals (Al, Zn), oxygen and elements from the interlayer anions (N, Cl, S, V) is consistent with the morphology of the coating for all the samples (Fig. 1). The distribution of N, Cl, S, and V tends to repeat the flake pattern of LDH coating, which confirms the presence of the anions in the LDH composition. The integral EDS spectra (Fig. 2) present a cumulative signal acquired for the corresponding area of each sample.

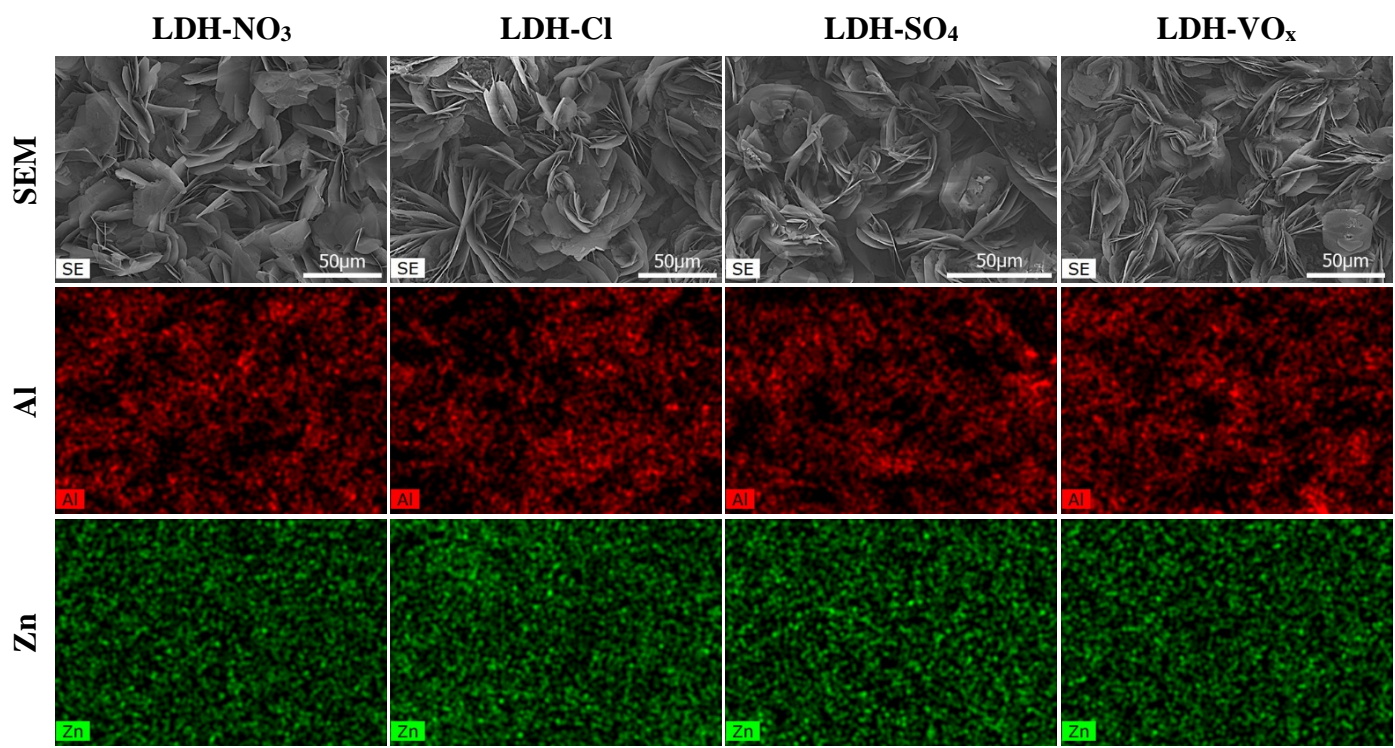

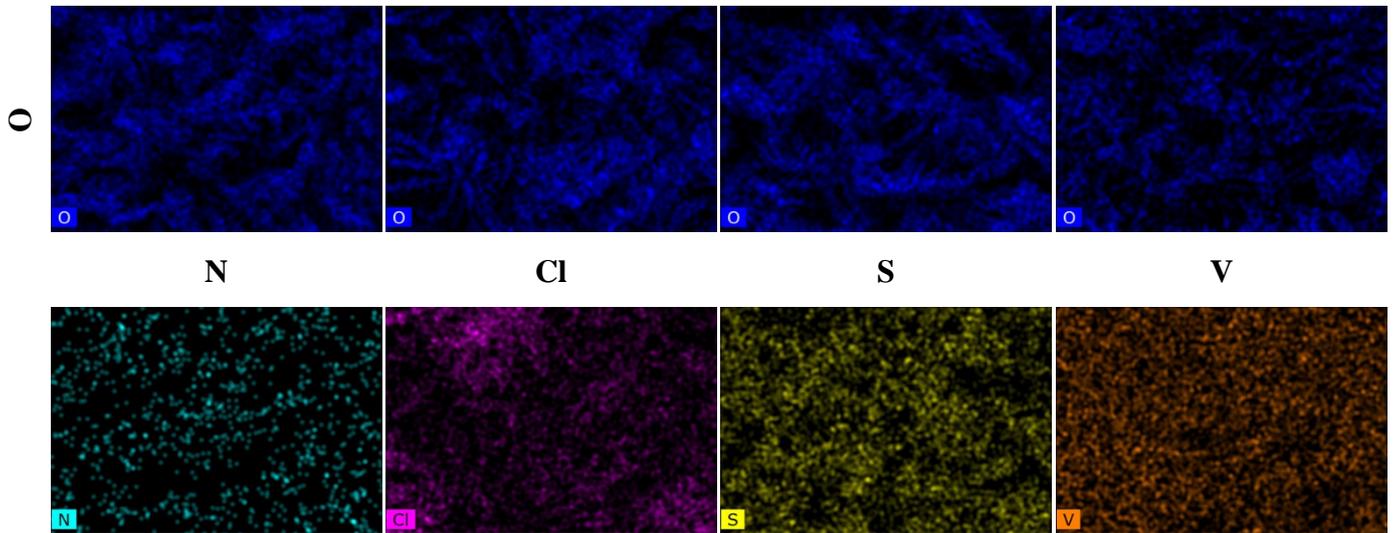

*Figure 1. SEM-images and EDS maps of LDHs with $NO_3^-$, $Cl^-$, $SO_4^{2-}$, $VO_x^{y-}$.*

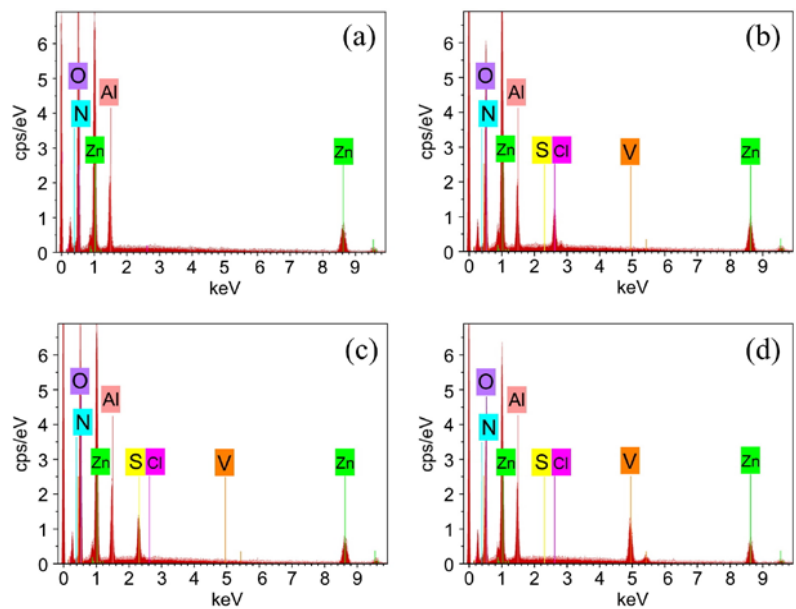

*Figure 2. EDS spectra of LDHs with a) $NO_3^-$, b) $Cl^-$, c) $SO_4^{2-}$, d) $VO_x^{y-}$*

### 3.2 X-Ray diffraction analysis

At the beginning of the intercalation reaction the crystal structure of the initial LDH-$NO_3$ was found to be a single-phase LDH with a high crystallinity. It was successfully refined within the rhombohedral space group R-3m with unit cell parameters of $a = 3.0771(2)$ Å and $c = 26.862(9)$ Å. These parameters correspond to those established in [36] permitting to conclude that in the Zn-LDH-$NO_3$, the nitrate ion plane is oriented with an angle of ~ 70° to the mix-metal hydroxide layer.

The anion exchange kinetics was analyzed based on the *in situ* obtained diffraction patterns. The extent of the reaction was defined by the ratio $\alpha(t) = I_{hkl}(t)/I_{hkl}(max)$, where $I_{hkl}(t)$ is the integral intensity of ($hkl$) peak at time $t$, $I_{hkl}(max)$ is the maximum integral intensity of this peak [37]. For the reaction parameters calculations, we used the integral intensities of the basis reflections $I_{003}$ as the most intensive and reliable reflection [30].

### 3.2.1. Chloride intercalation into LDH-NO$_3$, grown on zinc substrate (Zn-LDH-NO$_3$ →Zn-LDH-Cl)

The process of crystal structure evolution during anion exchange can be seen in Fig. 3a. The appearance of new diffraction peaks indicates the emergence of new crystalline phase. As one can see from Fig.3b, the positions of the new 00$l$ peaks are shifting to higher angles while a larger NO$_3^-$ anion is replaced by a smaller Cl$^-$ anion. We note that at the same time the (110) peak didn't change its position, this indicates that the lattice planes perpendicular to the 00$l$ do not change. We speculate that the chemical composition of the hydroxide layers in LDH is not changing [38].

The time dependence of the intensities of the basic reflections $I_{003}$ is shown in Fig. 3c, which reveals the process of nitrate being replaced by chloride in the Zn-LDH. As one can see, the anion exchange process takes place in two stages with an initial formation of an intermediate crystalline phase and the subsequent formation of the new LDH phase. This was already observed previously for the intercalation processes in LDHs [37].

From the crystallographic analysis of the diffraction patterns obtained at different moments reported in [33], we see shift of 00$l$ peaks towards large angle values. More detailed analysis of changing 00$l$ peak positions allows to conclude that a quick release of the nitrate anions (black squares) is accompanied by a quick formation and a slower conversion of the intermediate crystalline phase (IP, red circles) into the final crystalline phase Zn-LDH-Cl (blue triangles).

A schematic representation of the anion exchange process occurring between the nitrate and chloride anions in the galleries is shown in Fig. 4. The left panel of Figure 4 (I) shows the schematic arrangement of nitrate anions between the hydroxide layers, and the interplanar distance $d_{003}$ = 8.95 Å is equal to the distance between the cation layers.

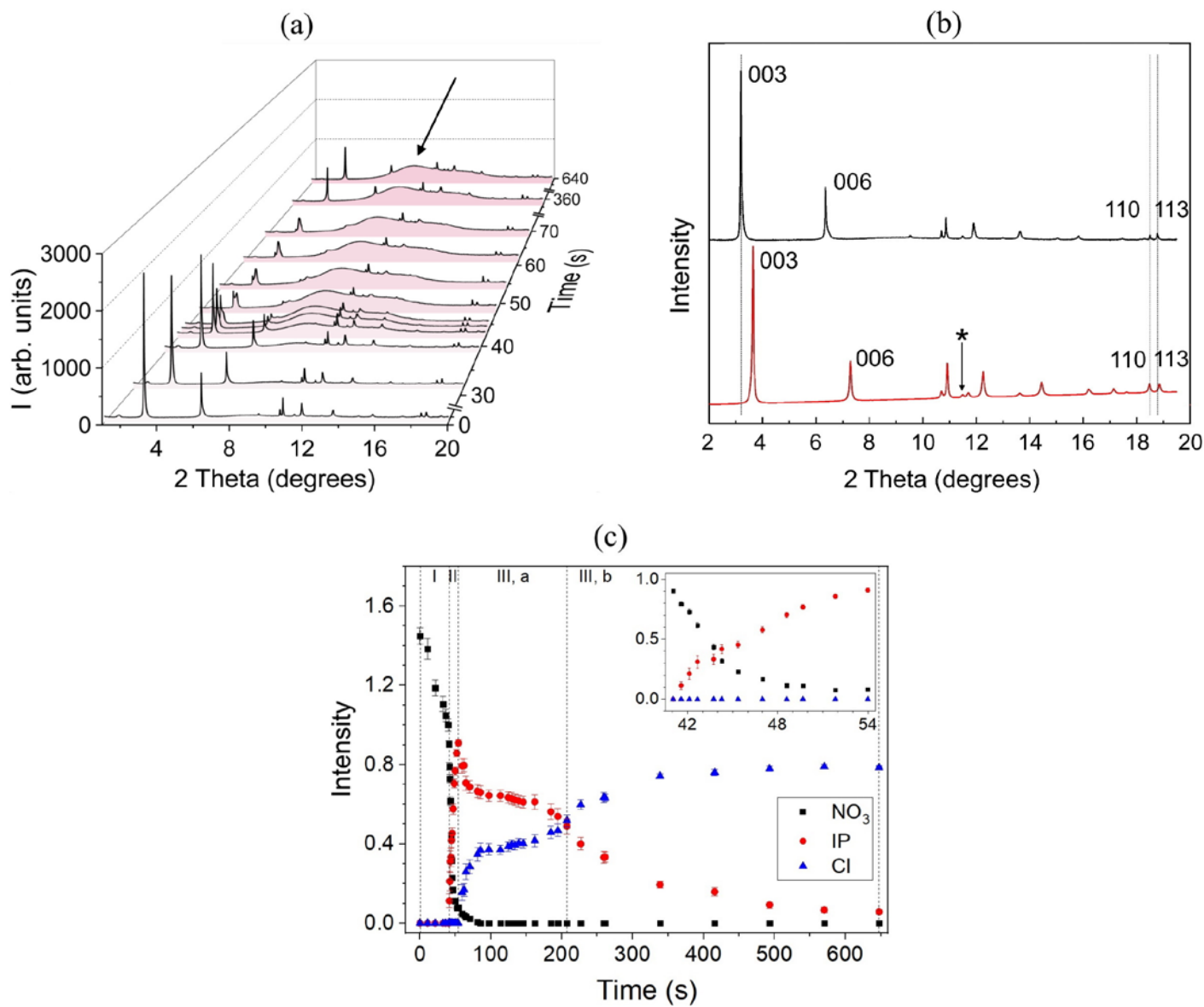

*Figure 3. a) Time evolution of XRD patterns during intercalation chloride anions into LDH-NO$_3$ (amorphous phase from water signal is noted by arrow); b) XRD patterns of initial LDH-NO$_3$ (black) and final LDH-Cl (red) phases. The amorphous phase was removed using AMORPH program. Non-LDH peak are denoted by asterisk (\*); c) Time evolution of integral intensity of 003 peak of the initial (LDH-NO$_3$, squares), intermediate (IP, circles) and final (LDH-Cl, triangles) phases. The broadened time scale from 33 to 55 seconds is shown at inset.*

The intermediate phase is shown in Figure 4 in the center panel (II). In this phase some galleries still contain anions NO$_3^-$ while another ones already are filled with Cl$^-$ anions. It can be represented as a crystal structure, where along the $c$ axis the galleries with remaining NO$_3^-$ anions, alternate with the galleries where the Cl$^-$ anions have already replaced the NO$_3^-$. The unit cell is doubled along $c$, the structure has the same symmetry R-3m with unit cell parameters a=3.150(2) Å and c=49.00(8) Å. In this case, the unit cell contains 6 rather than 3 layers along the $c$ axis. Since gallery heights of LDH-NO$_3$ and LDH-Cl are known, the value of $c$-parameter of IP gives a quantitative ratio of layers with NO$_3^-$ and Cl$^-$ of about 1:2 for IP (Fig. 4 II).

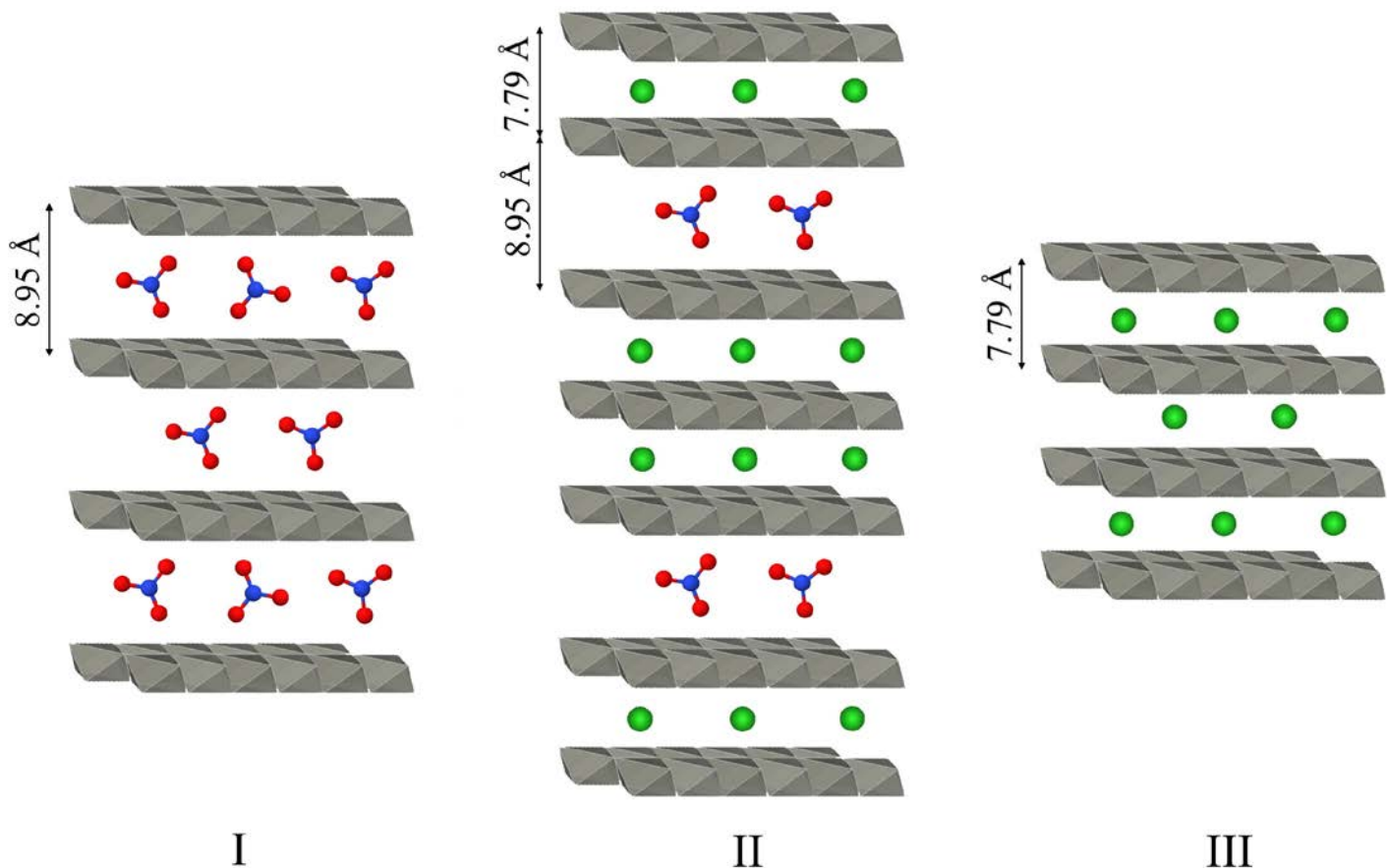

*Fig. 4. Schematic representation of the anion exchange process Zn-LDH-NO$_3$ →Zn-LDH-Cl:
I - The structure of the parent compound Zn-LDH-NO$_3$, II - intermediate phase, contains both nitrate anions and chloride, III - the final phase Zn-LDH-Cl.*

This leads to the point where that the Bragg peak near the position of (0 0 3) reflection of the initial crystal corresponds to *hkl* indices (0 0 6) of the intermediate phase. Finally, the right panel in Fig. 4 (III), represents the final crystalline phase, containing only chloride anions. The crystalline structure of the final phase of Zn-LDH-Cl can be still described by the hexagonal symmetry group R-3m.

There is also the possibility that during anion exchange reaction OH$^-$ can enter into the galleries, in addition to chloride. This can be seen either by changing of the symmetry (we would see extra peaks from LDH-OH), or by the worsening of the convergence in the crystal structure refinement within the R-3m group. Neither the former nor the latter was detected in the treatment of the diffraction patterns. However, we do not exclude presence of an insignificant amount OH$^-$ in the gallery with Cl$^-$. This doesn't complicate the current reasoning.

The entire period of anion exchange reaction can be divided into three time intervals: I - induction period, accompanied also by partial collapse of LDH-NO$_3$, II - active release of nitrate anions and the arrival of chloride anions (growth of the intermediate crystalline phase), IIIa,b – slow conversion of the intermediate phase via the formation and growth of the final phase, with different kinetic parameters, as shown in Table 3. In the inset of Fig. 3c, the interval II, which is very short, is represented at enlarged time scale.

For the analysis of the time dependencies of the reactions, we used the universal approach of Avrami-Erofeev (AE) [39–42]:

$$\alpha(t) = 1 - \exp\{-[k \cdot (t - t_0)]^m\} \qquad (1)$$

where $t_0$ is the time of induction period. In the AE model, a single reaction index $m$ is introduced, which combines the nucleation rate law with the growth mechanism of the nucleus. Additionally, there is the parameter $k$, which characterizes the reaction rate. Note that the rate varies if different phases appear during the progress of reaction development. In this regard, the AE model works most effectively at $0.15 < \alpha < 0.85$ and is not applicable for entirely diffusion-controlled processes [37]. The model is often used to explain a random nucleation and nuclei growth process. The determination of the reaction index $m$ allows to get an idea about the mechanisms involved in the exchange reaction. For the assessment of the applicability of the AE model, a preliminary estimation with Sharp-Hancock plots [43] was performed (Fig. 5), using the equation:

$$\ln(-\ln(1 - \alpha)) = m \cdot \ln k + m \cdot \ln(t - t_0) \qquad (2)$$

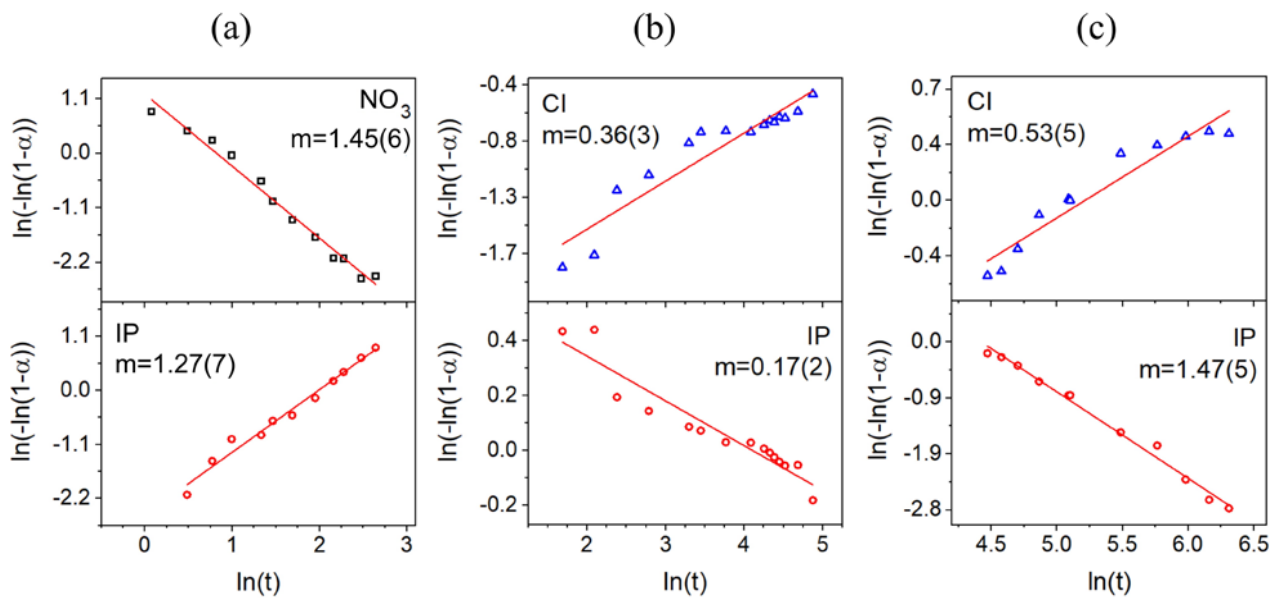

Fig. 5. Sharp-Hancock plots for Zn-LDH-NO$_3$ →Zn-LDH-Cl anion exchange: a) period II; b) period IIIa; c) period IIIb.

This equation can be considered as the logarithm of the AE equation. It should be noted that according to the work of S. F. Hulbert [44] a value of $m$ around 0.5 corresponds to a one-dimensional entirely diffusion-controlled reaction, for which the AE approach is not applicable. For the intermediate phase growth, the Sharp-Hancock plot gives a preliminary estimation of $m = 1.27(7)$, which corresponds to a two-dimensional diffusion-controlled reaction, decelerated by nucleation [37]. A similar value was obtained for the release of $NO_3^-$ anions from the parent compound, $m = 1.45(6)$. Using the AE equation (1) a least squares fitting for the intermediate crystalline phase during the reaction period II was done, which is displayed in Fig. 6a. In this way more accurate values of the reaction index $m$, the rate constant $k$ and active reaction start time $t_0$, both for the releasing LDH-NO$_3$ phase and for the new intermediate phase were obtained. These values are all summarized in Table 3.

The obtained $m$ values for the disappearing phase with $NO_3^-$ and for the forming IP with $NO_3^-/Cl^-$ are between 1 and 2, which confirms that the reaction mechanism is a two-dimensional diffusion-controlled one including a decelerated nucleation [37]. This means that the reaction rate is defined by the rate at which $Cl^-$ ions enter between the cation layers with decelerating effect of nucleation at the edges of the layers. Indeed, this seems to be consistent since LDH-NO$_3$ has a larger gallery height than that of LDH-Cl. The results further reveal that $NO_3^-$ starts to go out a little bit earlier and faster, than $Cl^-$ will replace it.

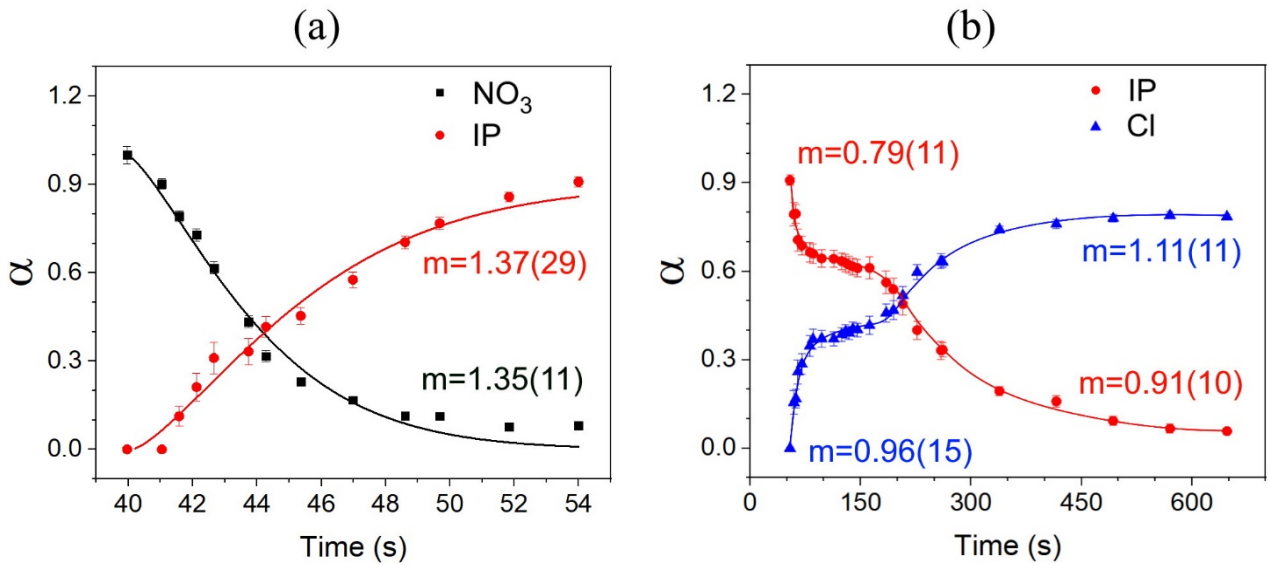

*Figure 6. Kinetic dependences of the degree of substitution of $NO_3^-$ by $Cl^-$ in Zn-LDH:*
*a) Interval II: release of $NO_3^-$ (black squares), IP formation (red circles); b) Interval III: IP decrease and formation of the final crystalline phase with $Cl^-$ intercalated (blue triangles). Solid lines are the fittings by the AE model.*

The final LDH-Cl phase growth takes place in two stages: IIIa and IIIb. For interval IIIa, the Sharp-Hancock plots give estimates of the reaction parameters of $m \sim 0.36$ and $0.17$ (Fig. 5b) and for interval IIIb of $m \sim 0.53$ and $1.47$ (Fig. 5c). As was mentioned, according to [44] for a one-dimensional entirely diffusion-controlled reaction ($m=0.5$) the AE approach is not applicable. In this case one can use the following expression for the reaction extent [37]

$$\alpha^2 = k \cdot (t - t_0) \qquad (3)$$

However, the analysis of the experimentally obtained dependences $\alpha(t)$ at IIIa, IIIb, carried out by the method of least squares using equations (1) and (3), with $R = \sqrt{R^2} = 0.99$ for (1) against $R = 0.75$ for (3) shows that the anion exchange process is much better described by the AE equation (1). Least mean squares fitting with the AE model provides *m*-values different from those obtained by the Sharp-Hancock approach (Fig. 6b), which exhibits rather great e.s.d.'s (estimated standard deviations) of integral intensities measurements. Thus, we conclude that in the time interval IIIa, a one-dimensional diffusion-controlled reaction takes place, with a decreasing rate of nucleation. A similar assumption is valid for IIIb interval, where Sharp-Hancock analysis gives *m*-values noticeably different from those ones obtained by least squares fitting to AE equation. The latter one gave *m*-indices, which lie in the range corresponding to one-dimensional diffusion-controlled reaction with decelerated nucleation rate.

It was shown [33] that the final phase reveals a good crystallinity and is described by R-3m space group with unit cell parameters $a = 3.081(2)$ Å, and $c = 23.359(5)$ Å. However, in the course of refinement of the atom positions, the best result is obtained for the coexistence of two crystalline phases with the same unit cell parameters [33], but with different positions of the Cl⁻ anions in the galleries which is summarized in Table 1. According this fact and since there are two similar kinetic processes with close parameters (IIIa and IIIb), it can be speculated that in these intervals the growth of the two crystalline phases in Zn-LDH-Cl occurs sequentially.

Table 1. Atomic positions of Cl⁻ and water oxygen in LDH galleries [33].

| Atom | Position | x | y | z | $R_p$ | $\chi^2$ |
|---|---|---|---|---|---|---|
| *Layer 1* | | | | | | |
| $O_w$ | 18h | 0.245(32) | 0.495(63) | 0.5 | 2.2 | 0.5 |
| Cl | 18h | 0.245(32) | 0.495(63) | 0.5 | | |
| *Layer 2* | | | | | | |
| $O_w$ | 18h | 0.881(5) | -0.881(5) | 0.5 | | |
| Cl | 18h | 0.881(5) | -0.881(5) | 0.5 | | |

*3.2.2. Sulphate intercalation into LDH-NO₃, grown on zinc substrate (Zn-LDH-NO₃ → Zn-LDH-SO₄)*

The refinement of the crystal structure of intercalated Zn-LDH-SO₄ reveals that it belongs to the symmetry group P-3 with parameters of $a = 5.3384(6)$ Å and $c = 11.1319(6)$ Å. This transformation is obviously connected with the substitution of the plane $NO_3^-$ anion by a 3D $SO_4^{2-}$ one, which cannot fit into a unit cell with a lattice constant $a \sim 3$ Å. The arrangement of SO₄-tetrahedra in the LDH gallery is represented in Fig. 7 according to the refined atomic positions, which are implicitly shown in Table 2.

Table 2. Refined atomic position in the unit cell of LDH-SO₄. $O_h$ and $O_w$ are related to OH-groups and water respectively, $O_1$ and $O_2$ are related to $SO_4^{2-}$.

| Atom | Position | x | y | z | $R_p$ | $\chi^2$ |
|---|---|---|---|---|---|---|

| Zn | 2d | 2/3 | 1/3 | 0 | | |
|---|---|---|---|---|---|---|
| Al | 1a | 0 | 0 | 0 | | |
| $O_h$ | 6g | 0 | 0.3333 | 0.061 (1) | 6.8 | 1.8 |
| S | 2c | 0 | 0 | 0.366 (1) | | |
| $O_1$ | | 0 | 0 | 0.461 (4) | | |
| $O_2$ | 6g | 0.14 (1) | 0.35 (3) | 0.335 (1) | | |
| $O_w$ | | 0.551 (8) | -0.202 (7) | 0.878 (2) | | |

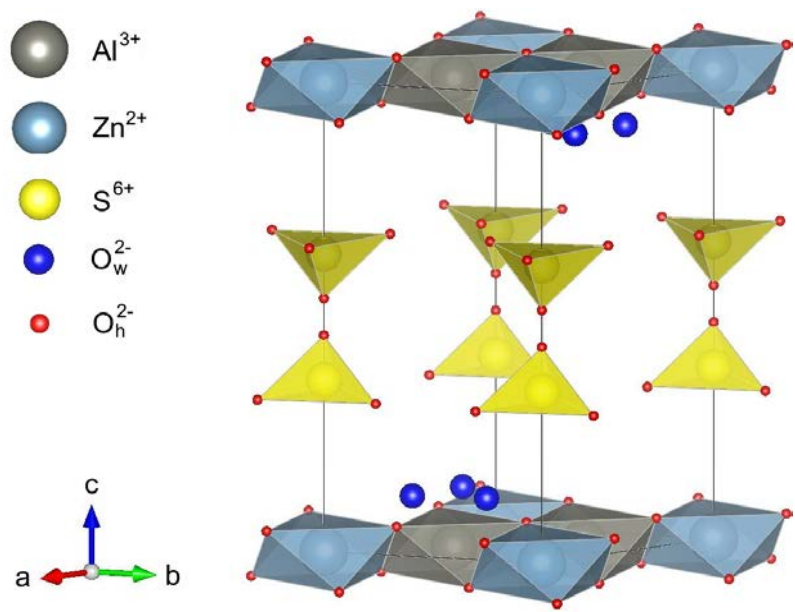

*Figure 7. Graphic representation of atomic positions in the unit cell of LDH-SO$_4$*

The time evolution of the diffraction pattern during anion exchange reaction Zn-LDH-NO$_3$ → Zn-LDH-SO$_4$ is shown in Fig. 8a. The shifts of 00$l$ and 110 peaks are displayed in Fig. 8b. In Fig. 8c the complete reaction is shown based on the intensity variation of the basal reflections. The black squares correspond to the collapse of the LDH-NO$_3$ phase, the red circles indicate the formation and final conversion of the intermediate crystalline phase and the blue triangles show the formation and growth of the final crystalline phase Zn-LDH-SO$_4$.

The sulphate anion exchange reaction takes place also via an intermediate phase, similar to the situation presented in Fig.4. In the present case the unit cell parameter *c* for the IP will be the sum of 11.11 Å and 8.95 Å. For the calculation of the reaction parameters, the integral intensities of the basis reflections $I_{003}$ of parent Zn-LDH-NO$_3$, $I_{002}$ for IP and $I_{001}$ for Zn-LDH-SO$_4$ were used.

The total duration of the reaction process could be divided into the following time intervals, similar to the previous case: I - induction period, accompanied also by a partial collapse of LDH-NO$_3$, II - active release of nitrate and the arrival of sulphate anions - growth of the intermediate crystalline phase (IP), interval III – disappearing of intermediate phase, and formation and growth of the final phase. In contrast to the reaction with chloride, the process is not a perfect anion exchange reaction, since the intermediate phase is not converted completely, as it can be clearly seen by comparing Fig. 3c and Fig. 8c.

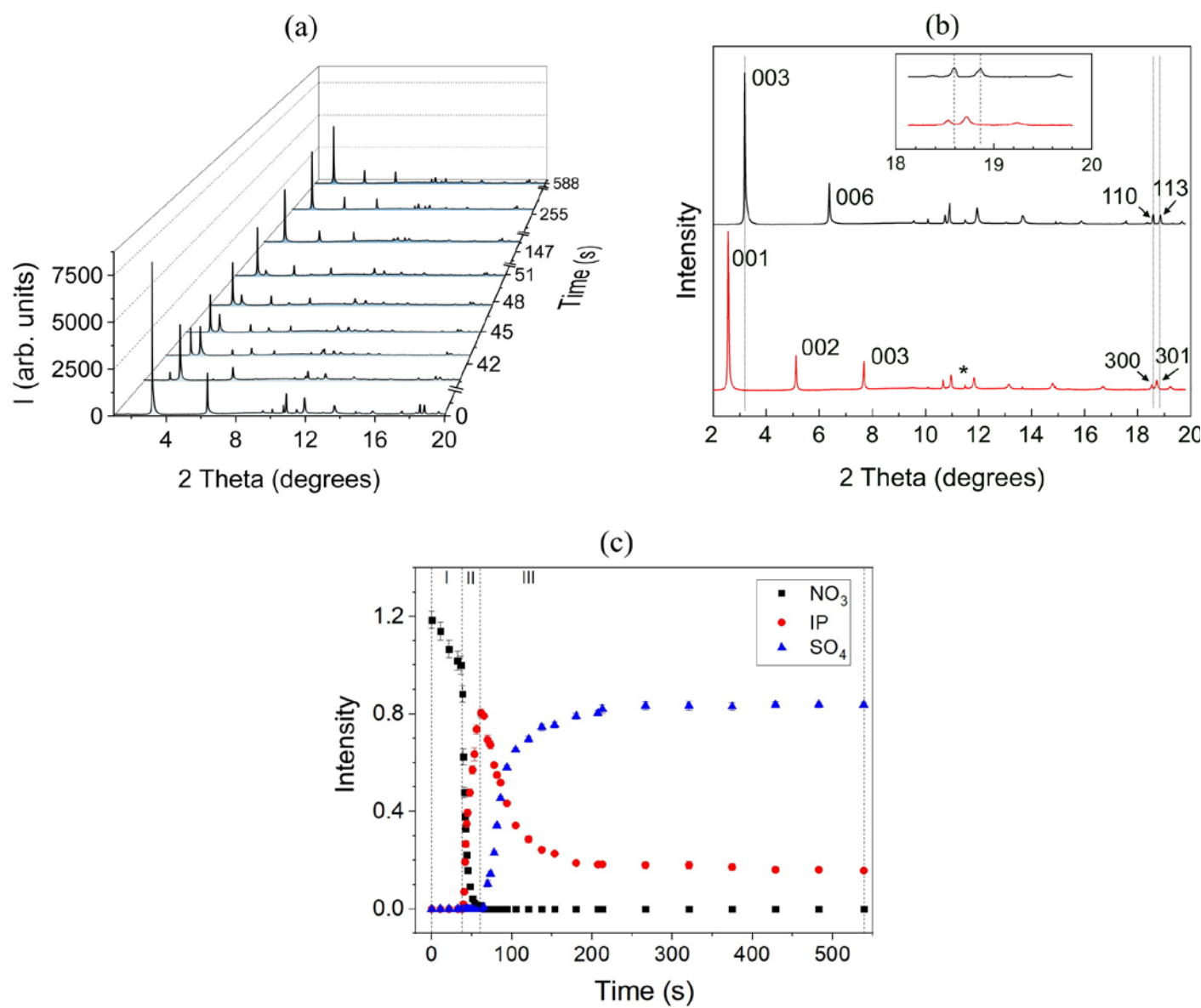

Fig. 8. a) Time evolution of XRD patterns during $NO_3^-$–$SO_4^{2-}$ exchange; b) XRD patterns of initial LDH-NO$_3$ (black) and final LDH-SO$_4$ (red) phases. Non-LDH peaks are denoted by asterisk (*). Inset shows the broadened angle scale from 18 to 20 degrees. c) Time evolution of integral intensity of 003 peak: decay of the initial phase Zn-LDH-SO$_4$ (squares), formation and decrease of IP (circles) and formation of Zn-LDH-SO$_4$ (triangles).

For the analysis of the intercalation kinetics for Zn-LDH-NO$_3$ → Zn-LDH-SO$_4$, the same procedure was followed as in section 3.2.1. In the first step, the preliminary estimations with the Sharp-Hancock plot were performed (Fig. 9) to verify the applicability of the AE model for the analysis of the dependences of kinetics in this case. According to these plots for interval II the reaction parameter $m$ is around 1.99 for the release of nitrate anions, while $m$ is 1.75 for the growth of the intermediate phase. Similar to the exchange reaction for chloride, the obtained values correspond to a two-dimensional diffusion-controlled reaction decelerated by nucleation. For the reaction interval III the $m$-index was ~ 0.8 for the final phase growth and for the decreasing intermediate phase *m* was ~ 0.58. The latter values are once more close to the critical value of 0.5, corresponding either to one-dimensional diffusion-controlled reaction with deceleratory nucleation rate or to one-dimensional entirely diffusion controlled reaction. Therefore, the fit of this dependence was done with both equations (1), and (3). The parameter *R* was ~ 0.99 for (1), and *R* ~ 0.71 for (3), which clearly indicates that the reaction can be described by the AE model (1).

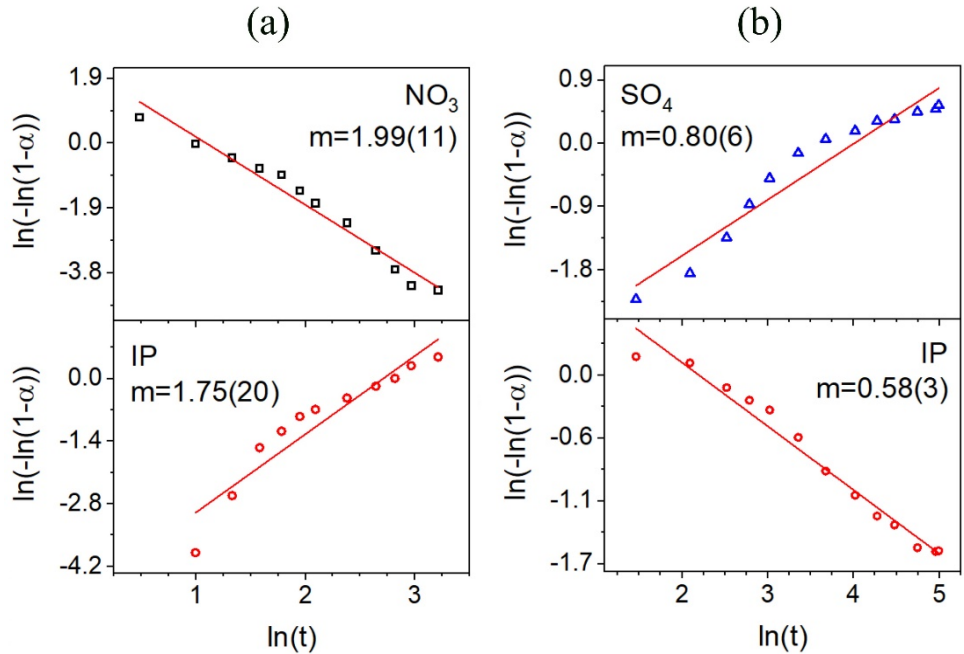

*Figure 9. Sharp-Hancock plots for Zn-LDH-NO$_3$ →Zn-LDH-SO$_4$ exchange: a) period II; b) period III.*

The fitting curves are presented in Fig. 10. The exact kinetic parameters obtained in this way are presented in Table 3. Similar to the case of Zn-LDH-NO$_3$ →Zn-LDH-Cl exchange, the final *m*-values differ from the ones, obtained preliminary by the Sharp-Hancock plots due to rather great e.s.d.'s of integral intensities measurements. Here as well, the final *m*-values for period III correspond to a one-dimensional diffusion-controlled reaction with effect of deceleratory nucleation. As it was noted above the intermediate phase remains (see Fig. 10b). The quantity of IP in the sample is still about 20% while the final phase reaches about 80%.

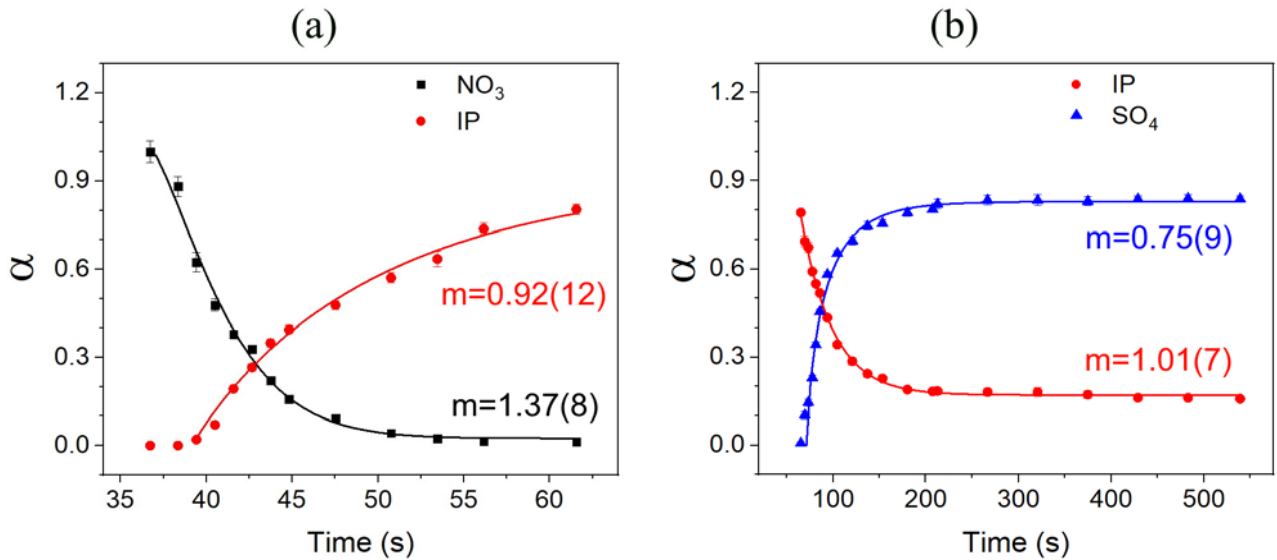

*Figure 10. Kinetic dependences of the degree of substitution of $NO_3^-$ by $SO_4^{2-}$ in Zn-LDH: a) Interval II: release of $NO_3^-$ (black squares), IP growth (red circles); b) Interval III: IP conversion and formation of the final crystalline phase with $SO_4^{2-}$ intercalated (blue triangles). Solid lines are the fittings by AE model.*

### 3.2.3. Vanadate intercalation into LDH-NO3, grown on zinc substrate (Zn-LDH-NO3 → Zn-LDH-VOx)

Time evolution of the diffraction pattern and shifts of basal peaks are shown in Fig. 11a,b. The basal peak 00$l$ positions shift to smaller angles which is associated to an increase of galleries heights which is related to incoming vanadate anions. It is visible from the final diffraction pattern that the basal peaks are doubled, thus indicating the coexistence of two new phases, which we designate as LDH-VO$_x$(1) and LDH-VO$_x$(2). The quality of the pattern does not permit the identification of the vanadate speciation in the interlayers in these phases. Only the basal spacing of the intercalated compounds: $d = 9.729(2)$ Å for LDH-VO$_x$(1) and $d = 10.348(3)$ Å for LDH-VO$_x$(2) can be obtained. The anion exchange reaction takes place in a solution with a pH of about 7–8. This could result in several types of intercalated polyvanadates depending also on the vanadium concentration. Anions, such as $V_4O_{12}^{4-}$ along with $V_2O_7^{4-}$ are the species resulting from the polymerization under these conditions [45, 46].

The gallery heights for the flat arrangement of these anions in the LDHs interlayers are close to each other and are equal to 4.7 Å for $V_4O_{12}^{4-}$ and to 5.0 Å for $V_2O_7^{4-}$ [45, 47]. The gallery heights obtained for our measurements are equal to 5.045(3) Å for the phase LDH-VO$_x$(1), and 5.624(2) Å for the phase LDH-VO$_x$(2). This confirms the assumption, that two vanadate species $V_4O_{12}^{4-}$ and $V_2O_7^{4-}$ are present in LDH matrices. A similar finding was reported in [45], where the coexistence of two layered phases containing two different vanadate anions was noted.

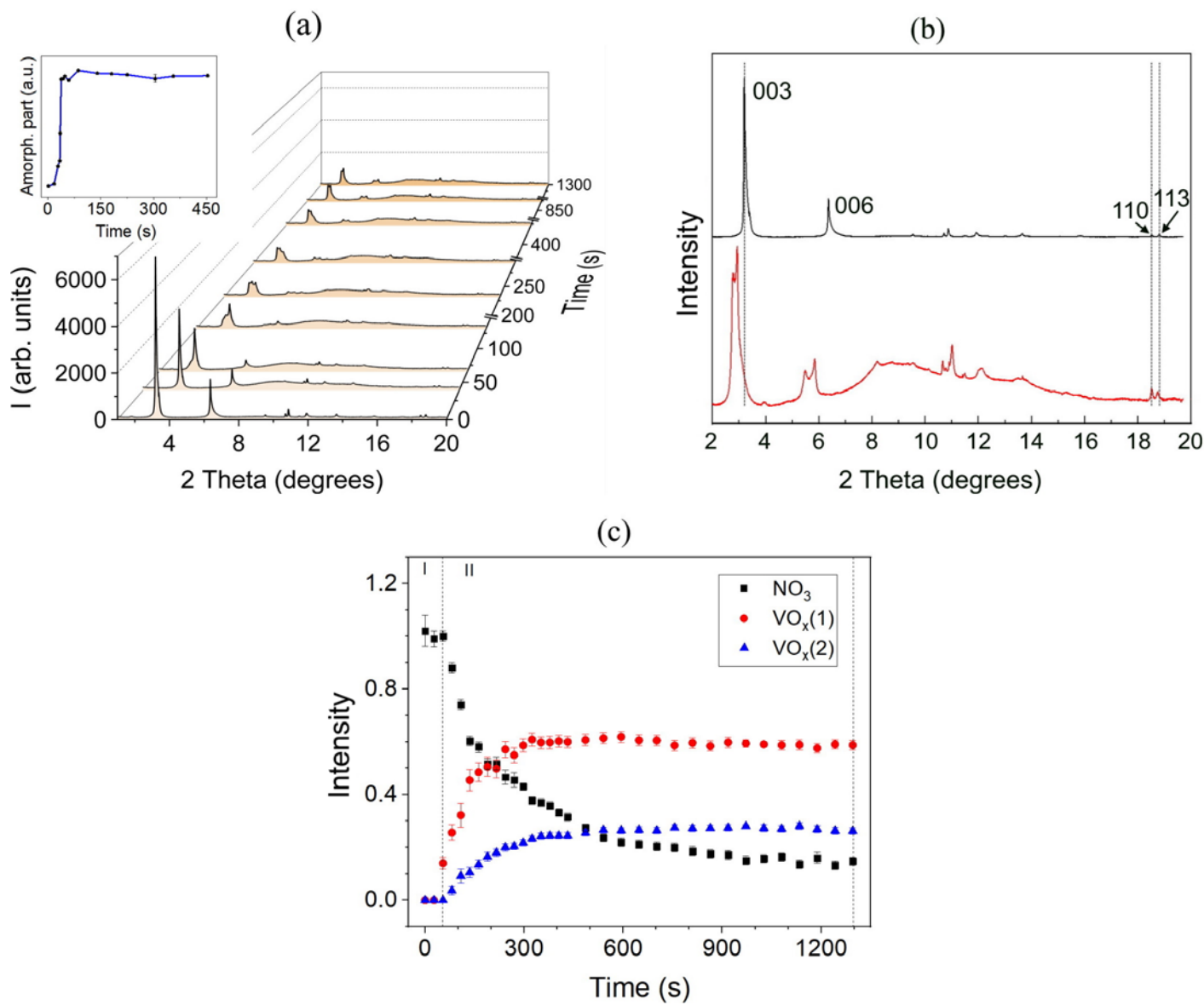

*Figure 11. a) Time evolution of XRD patterns during $NO_3^-$–$VO_x^{y-}$ exchange; b) XRD patterns of initial LDH-NO₃ (black) and final LDH-VO$_x$ (red) phases. c) Time evolution of integral intensity of 003 peak of the initial phase Zn-LDH-NO₃ (squares), and formation of the two final intercalation products LDH-VO$_x$(1) (circles) and for Zn-LDH-VO$_x$(2) (triangles).*

In contrast to the previous two reactions, the anion exchange is going without the intermediate phase. The process can be divided into 2 stages: the first one is the induction period and the second one is the release of nitrates and the growth of the new crystal phases. After the induction period, which lasts a similar time interval as for the exchange reactions of $NO_3^- \rightarrow Cl^-$, or $NO_3^- \rightarrow SO_4^{2-}$, the vanadate phases begin to grow directly. The time evolution of both vanadate phases is shown in Fig. 11c by red circles and blue triangles, respectively. Furthermore, the nitrate anions did not leave the LDH matrix completely which is indicated by the fact that reflections from parent LDH-NO₃ do not disappear until the end of intercalation process.

Similar to the above described analysis the first step of evaluation is the use of the Sharp-Hancock equation for the analysis of the time dependent change of the Bragg peaks (see Fig. 12). The results are $m \sim 0.63$ for the release stage of $NO_3^-$ anions and $m \sim 0.32$ and $m \sim 0.55$, for the growth of the two new phases, respectively.

In the case of this one-stage intercalation process, one should expect a two-dimensional entirely diffusion-controlled reaction since these *m*-values are close to 0.5. The equation which describes such a process is [37]:

$$(1 - \alpha) \ln(1 - \alpha) + \alpha = k \cdot (t - t_0) \tag{4}$$

The application of formula (4) gave the parameter of $R \sim 0.68$. When the AE equation (1) was used a value of $R \sim 0.98$ was obtained. Thus, the final fitting was done using AE equation (1) and calculated curves are shown in Fig. 13.

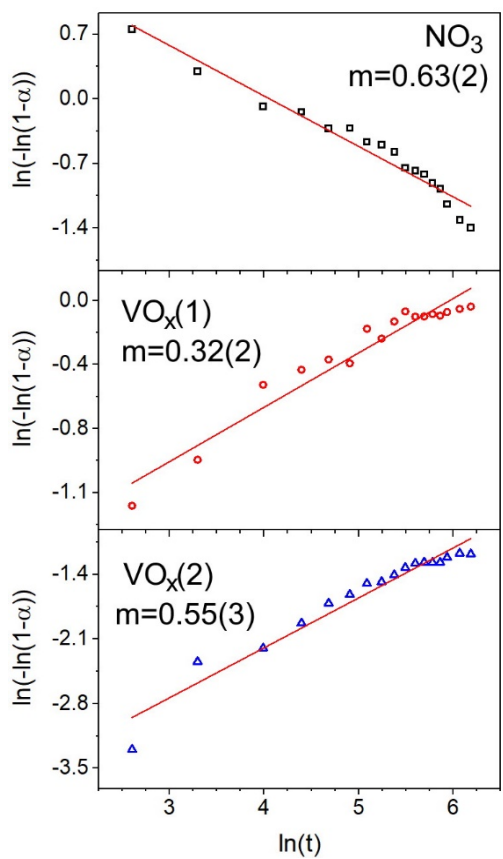

*Figure 12. Sharp-Hancock plots for Zn-LDH-NO$_3$ →Zn-LDH-VO$_x$ exchange (period II).*

The kinetic parameters for the growth of both phases with $VO_x^{y-}$ are given in Table 3. The obtained reaction indices are: $m = 1.18(15)$ for $VO_x^{y-}(1)$ and $m = 0.99(7)$ for $VO_x^{y-}(2)$. This means that the formation and growth of the final crystal phases with $VO_x^{y-}$ is characterized by two-dimensional diffusion-controlled reaction following instantaneous nucleation. The obtained value for *m* of 1.00(4) for the release of $NO_3^-$ suggests that the release of the nitrate anions can be regarded as a two-dimensional diffusion controlled reaction with zero nucleation rate.

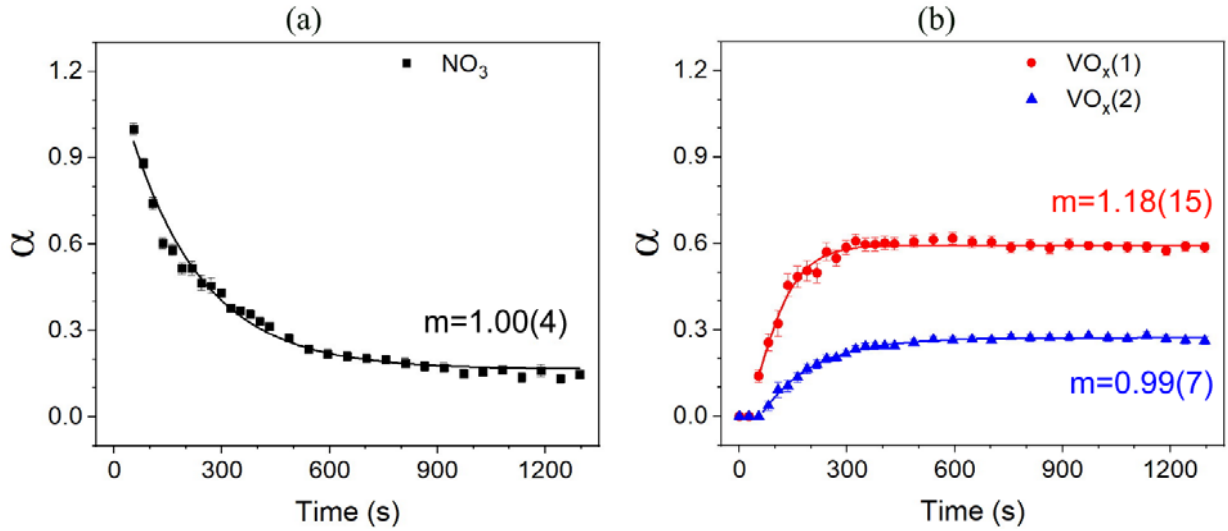

*Figure 13. Kinetic dependences of the degree of substitution of $NO_3^-$ by $VO_x^{y-}$ in Zn-LDH: a) release of $NO_3^-$ (black squares); b) formation of two crystalline phases with $VO_x^{y-}$ intercalated (red circles and blue triangles). Solid lines are the fittings by AE model.*

Table 3. Kinetic parameters for intercalation reactions Zn-LDH-NO$_3$ → Zn-LDH-Cl, Zn-LDH-NO$_3$ → Zn-LDH-SO$_4$ and Zn-LDH-NO$_3$ → Zn-LDH-VO$_x$. The units of measurement of the parameters k and t are s$^{-1}$ and s, respectively.

|  | $NO_3^- \to Cl^-$ | $NO_3^- \to SO_4^{2-}$ | $NO_3^- \to VO_x^{y-}(1)$ | $NO_3^- \to VO_x^{y-}(2)$ |
|---|---|---|---|---|
| LDH-NO$_3$ | $t_0$ = 39.9 (3.1) | $t_0$ = 36.6 (1.2) | $t_0$ = 32.4 (5.8) | |
|  | m = 1.35 (11) | m = 1.37 (9) | m = 1.00 (4) | |
|  | k = 0.22 (1) | k = 0.20 (1) | k = 0.0049 (3) | |
| IP(II) | $t_0$ = 40.1 (2.8) | $t_0$ = 38.9 (2.4) | – | – |
|  | m = 1.37 (29) | m = 0.92 (12) | | |
|  | k = 0.16 (3) | k = 0.09 (2) | | |
| IP(III,a) | $t_0$ = 53.5 (2.2) | $t_0$ = 65.2 (1.2) | – | – |
|  | m = 0.79 (11) | m = 1.01 (7) | | |
|  | k = 0.09 (1) | k = 0.029 (1) | | |
| IP(III,b) | $t_0$ = 179.6 (3.7) | – | | |
|  | m = 0.91 (10) | | | |
|  | k = 0.008 (1) | | | |
| Final crystalline phase (III, a) | $t_0$ = 59.3 (3.1) | $t_0$ = 71.1 (1.4) | $t_0$ = 48.9 (5.1) | $t_0$ = 66.0 (5.6) |
|  | m = 0.96 (15) | m = 0.75 (9) | m = 1.18 (15) | m = 0.99 (7) |
|  | k = 0.07 (1) | k = 0.049 (5) | k = 0.012 (1) | k = 0.0069 (4) |
| Final crystalline phase (III, b) | $t_0$ = 184.3 (2.0) | – | – | – |
|  | m = 1.11 (11) | | | |
|  | k = 0.011 (1) | | | |

## 3.3 Final Remarks

As one can see from the analysis of the first two anion exchange reactions, where $NO_3^-$ anions are replaced by $Cl^-$ and $SO_4^{2-}$ ones, the rate of guest diffusion within individual layers is large compared with the rate of nucleation (see Sec. 3.2.1 and 3.2.2, Table 3). It means that an appreciable number of layers is filled before all potential nucleation sites are saturated. The decreasing number of nucleation sites decelerate the total rate of the reaction in this case, and the rate of saturation of nucleation sites is comparable with the rate of diffusion. The intercalation process Zn-LDH-$NO_3$ → Zn-LDH-Cl has the highest rate (see Table 3). The high rate of the guest diffusion in this reaction is expected since a plane anion such as $NO_3^-$, placed almost perpendicular to cation layers, is exchanged by a single $Cl^-$ anion. The gallery height for Zn-LDH-$NO_3$ is $h = 4.193(4)$ Å which is considerably larger than $h = 3.076(2)$ Å for Zn-LDH-Cl [33]. This provides auspicious conditions for the accommodation of $Cl^-$ anions within cation layers. The gallery height $h$ was calculated by substracting the thickness of hydroxide layer (~ 4.71 Å [48]) from interplanar distance $d_{003}$ ($d_{001}$ for LDH-$SO_4$).

For the reaction Zn-LDH-$NO_3$ → Zn-LDH-$SO_4$ the obtained reaction rates are also comparatively high. The gallery height for Zn-LDH-$NO_3$ is $h = 4.193(4)$ Å and the height of the sulphate $SO_4^{2-}$ tetrahedra is $h_t \sim 2.6$ Å. Obviously such a small anion can intercalate in the larger space sufficiently fast, and the rate should be more or less similar as it was for chloride intercalation. Nevertheless, the rate here is noticeably lower. Furthermore, the gallery height $h = 6.422(1)$ Å of the final phase Zn-LDH-$SO_4$ is larger than that of Zn-LDH-$NO_3$ suggesting that other factors like electrostatic interactions have to be accounted. It is likely that electrostatic interactions pull the adjacent $SO_4^{2-}$ tetrahedral anions to the opposite cation layers thus increasing the interlayer distance. This process could be also considered as a hindrance to complete the reaction of anion exchange, thus the intermediate phase remains in the sample.

The reaction Zn-LDH-$NO_3$ → Zn-LDH-$VO_x$ has a low reaction rate and, thus, requires much more time for the anion exchange than for the chloride and sulphate ones (see Table 3). The LDH-$VO_x$(2) phase has the lowest growth rate parameter $k = 0.0069(4)$. The deceleratory effect of nucleation, which was identified for reactions with higher rates Zn-LDH-$NO_3$ → Zn-LDH-$SO_4$ and Zn-LDH-$NO_3$ → Zn-LDH-Cl, does not manifest itself on the time scale of Zn-LDH-$NO_3$ → Zn-LDH-$VO_x$ reaction. From this we can deduce that the nucleation sites are completely saturated much faster than diffusion goes on; so, the nucleation could be regarded as simultaneous and could be the reason for the absence of an intermediate phase. Because of the low rate of transport of guest anions between the cation layers, they have enough time to start filling all the interlayer galleries before a significant number of them are completely filled. The low diffusion rate is obviously connected with the large size of the anions which are presumably $V_4O_{12}^{4-}$ and $V_2O_7^{4-}$. The influence of anion charge on reaction rate requires additional studies.

From these results one can deduce that the rate of replacing host nitrate anions by guest ones can be represented as $Cl^- > SO_4^{2-} > VO_x^{y-}$.

## Conclusions

Synchrotron *in situ* studies were used to obtain kinetic parameters of anion exchange reactions Zn-LDH-NO$_3$ → Zn-LDH-SO$_4$, Zn-LDH-NO$_3$ → Zn-LDH-Cl and Zn-LDH-NO$_3$ → Zn-LDH-VO$_x$. It was demonstrated that intercalation for the processes NO$_3^-$ → SO$_4^{2-}$ and NO$_3^-$ → Cl$^-$, is faster and both reactions occur in two stages. The first stage is characterized by two-dimensional diffusion-controlled reaction following deceleratory nucleation, and then in the second stage by a one-dimensional diffusion-controlled reaction also with a deceleratory nucleation effect. In the case of intercalation Zn-LDH-NO$_3$ → Zn-LDH-Cl the anion exchange NO$_3^-$ by Cl$^-$ ends by complete release of the host anions. On contrary, for Zn-LDH-NO$_3$ → Zn-LDH-SO$_4$ this process does end with an incomplete release of NO$_3^-$.

The intercalation process Zn-LDH-NO$_3$ → Zn-LDH-VO$_x$ takes much more time than the previous ones and is characterized by a one stage two-dimensional reaction with instantaneous nucleation. At the end of the reaction two new LDH structures are formed with different polyvanadate species intercalated, presumably V$_4$O$_{12}^{4-}$ and V$_2$O$_7^{4-}$. The release of NO$_3^-$ anion is also incomplete in this case.

The exchange reaction rate for chloride, sulphate and vanadate anions during anion exchange in LDH-NO$_3$ decreases according to the sequence Cl$^-$ > SO$_4^{2-}$ > VO$_x^{y-}$.

## Acknowledgements

We acknowledge DESY (Hamburg, Germany), a member of the Helmholtz Association HGF, for the provision of experimental facilities. Parts of this research were carried out at **PETRA III** under the proposal number **I-20170366**. We would like to thank Dr. Sergey Volkov for assistance in using the **diffractometer in P08 high-resolution diffraction beamline.**

M.H.I and A.C.B are grateful for the financial support of the German-Russian Interdisciplinary Science Center (G-RISC) in form of a travel grant (T-2018b-1 and T-2018a-3 respectively) that enables them to perform a scientific exchange and complete this work.

K.Y. thanks financial support of a researcher grant (IF/01284/2015) and both K.Y. and M.G.S.F thank the project CICECO-Aveiro Institute of Materials, UIDB/50011/2020 & UIDP/50011/2020, financed by national funds through the FCT/MEC and when appropriate co-financed by FEDER under the PT2020 Partnership Agreement.

**References**

1. Allmann, R. The crystal structure of pyroaurite. *Acta Crystallogr. Sect. B: Struct. Crystallogr. Cryst. Chem.,* **1968,** 24, 972–977, DOI: 10.1107/S0567740868003511.

2. Li, B.; He, J.; Evans, D.G.; Duan, X. Inorganic layered double hydroxides as a drug delivery system—intercalation and in vitro release of fenbufen. *Appl. Clay Sci.*, **2004**, *27*, 199–207, DOI: 10.1016/j.clay.2004.07.002.

3. Yan, K.; Liu, Y.; Lu, Y.; Chai, J.; Sun, L. Catalytic application of layered double hydroxide-derived catalysts for the conversion of biomass-derived molecules. *Catal. Sci. Technol.*, **2017**, *7*, 1622–1645, DOI: 10.1039/C7CY00274B.

4. Goh, K.-H.; Lim, T.-T.; Dong, Z. Application of layered double hydroxides for removal of oxyanions: a review. *Water Res.*, **2008**, *42,* 1343–1368, DOI: 10.1016/j.watres.2007.10.043.

5. Zubair, M.; Daud, M.; McKay, G.; Shehzad, F.; Al-Harthibe, M. A. Recent progress in layered double hydroxides (LDH)-containing hybrids as adsorbents for water remediation. *Appl. Clay Sci.*, **2017**, *143*, 279–292, DOI: 10.1016/j.clay.2017.04.002c.

6. Taylor, H. F. W. Segregation and cation-ordering in sjögrenite and pyroaurite. *Mag.*, **1969**, *37*, 338–342, DOI: 10.1180/minmag.1969.037.287.04.

7. Birgul, Z. K.; Ahmet, A. Layered double hydroxides – multifunctional nanomaterials. *Chem. Pap.,* **2012**, *66*, 1–10, DOI: 10.2478/s11696-011-0100-8.

8. Sels, B.; Vos, D.; Buntinx, M.; Pierard F.; Mesmaeker A. K.-D.; Jacobs P. Layered double hydroxides exchanged with tungstate as biomimetic catalysts for mild oxidative bromination. *Nature* **1999**, *400*, 855–857, DOI:10.1038/23674.

9. Kovanda, F.; Kovacsova, E.; Kolousek, D. Removal of anions from solution by calcined hydrotalcite and regeneration of used sorbent in repeated calcination–rehydration–anion exchange processes. *Collect. Czech. Chem. Commun.*, **1999**, *64*, 1517–1528, DOI: 10.1135/cccc19991517.

10. Xu, Z. P.; Zeng, H. C. Ionic interactions in crystallite growth of CoMgAl-hydrotalcite-like compounds. *Chem. Mater.,* **2001**, *13*, 4555-4563, DOI: 10.1021/cm010222b.

11. Carlino, S. The intercalation of carboxylic acids into layered double hydroxides: a critical evaluation and review of the different methods. *Solid State Ionics*, **1997***, 98*, 73-84, DOI: 10.1016/S0167-2738(96)00619-4.

12. Aisawa, S.; Takahashi, S.; Ogasawara, W.; Umetsu, Y.; Narita, E. Direct intercalation of amino acids into layered double hydroxides by coprecipitation. *Journal of Solid State Chemistry*, **2001**, *162*, 52-62, DOI: 10.1006/jssc.2001.9340.

13. Choy, J.-H.; Kwak, S.-Y.; Park, J.-S.; Jeong, Y.-J. Cellular uptake behavior of [γ-$^{32}$P] labeled ATP–LDH nanohybrids. *J. Mater. Chem.*, **2001**, *11*, 1671-1674, DOI: 10.1039/B008680K.

14. Nyambo, C.; Songtipya, P.; Manias, E.; Jimenez-Gascoc M. M.; Wilkie, C. A. Effect of MgAl-layered double hydroxide exchanged with linear alkyl carboxylates on fire-retardancy of PMMA and PS. *J. Mater. Chem.*, **2008**, *18*, 4827–4838, DOI: 10.1039/B806531D.

15. Alcântara, A. C. S.; Aranda, P.; Darder, M.; Ruiz-Hitzky, E. Bionanocomposites based on alginate–zein/layered double hydroxide materials as drug delivery systems. *J. Mater. Chem.*, **2010** 20, 9495–9504, DOI: 10.1039/C0JM01211D.

16. Chen, H.; Hu, L.; Chen, M.; Yan, Y.; Wu, L. Nickel–Cobalt Layered Double Hydroxide Nanosheets for High-performance Supercapacitor Electrode Materials. *Adv. Funct. Mater.*, **2014**, *24*, 934–942, DOI: 10.1002/adfm.201301747.

17. Buchheit, R. G.; Guan, H.; Mahajanam, S.; Wong, F. Active corrosion protection and corrosion sensing in chromate-free organic coatings. *Prog. Org. Coat.*, **2003**, *47*, 174–182, DOI: 10.1016/j.porgcoat.2003.08.003.

18. Zheludkevich, M.L.; Poznyak, S.K.; Rodrigues, L.M.; Raps, D.; Hack, T.; Dick, L.F.; Nunes, T.; Ferreira, M.G.S. Active protection coatings with layered double hydroxide nanocontainers of corrosion inhibitor. *Corros. Sci.*, **2010**, *52,* 602–611, DOI: 10.1016/j.corsci.2009.10.020.

19. Scarpellini, D.; Falconi, C.; Gaudio, P.; Mattoccia, A.; Medaglia, P.G.; Orsini, A.; Pizzoferrato, R.; Richetta, M. Morphology of Zn/Al layered double hydroxide nanosheets grown onto aluminum thin films. *Microelectron. Eng.*, **2014,** *126*, 129–133, DOI: 10.1016/j.mee.2014.07.007.

20. Hang, T.; Truc, T.; Duong, N.; Vu, P.; Hoang, T. Preparation and characterization of nanocontainers of corrosion inhibitor based on layered double hydroxides. *Appl. Clay Sci.* **2012**, *67*, 18–25, DOI: 10.1016/j.clay.2012.07.004.

21. Li, Y.; Li, S.; Zhang, Y.; Yu, M.; Liu, J. Enhanced protective Zn–Al layered double hydroxide film fabricated on anodized 2198 aluminum alloy, *J. Alloys Compd.*, **2015**, *630,* 29–36, DOI: 10.1016/j.jallcom.2014.12.176.

22. Zhang, F.; Liu, Z.-G.; Zeng, R.-C.; Li, S.-Q.; Cui, H.-Z.; Song, L.; Hanc E.-H. Corrosion resistance of Mg–Al-LDH coating on magnesium alloy AZ31. *Surf. Coat. Tech.*, **2014**, *258*, 1152-1158, DOI: 10.1016/j.surfcoat.2014.07.017.

23. Mikhailau, A.; Maltanava, H.; Poznyak, S. K.; Salak, A. N.; Zheludkevich, M. L.; Yasakau, K. A.; Ferreira, M. G. S. One-step synthesis and growth mechanism of nitrate intercalated ZnAl LDH conversion coatings on zinc. *Chem. Commun.*, **2019**, *55*, 6878-6881, DOI: 10.1039/C9CC02571E.

24. Kuznetsov, B.; Serdechnova, M.; Tedim, J.; Starykevich, M.; Kallip, S.; Oliveira, M. P.; Hack, T.; Nixon, S.; Ferreira, M. G. S.; Zheludkevich, M. L. Sealing of tartaric sulphuric (TSA) anodized AA2024 with nanostructured LDH layers, *RSC Adv.*, **2016**, *6*, 13942-13952, DOI: 10.1039/C5RA27286F.

25. Hoshino, K.; Furuya, S.; Buchheit, R. G. Effect of $NO_3^-$ Intercalation on Corrosion Resistance of Conversion Coated Zn-Al-$CO_3$ LDHs on Electrogalvanized Steel. *J. Electrochem. Soc.*, **2018**, *165,* C461-C468, DOI: 10.1149/2.0091809jes.

26. Tedim, J.; Kuznetsova, A.; Salak, A. N.; Montemor, F.; Snihirova, D.; Pilz, M.; Zheludkevich, M. L.; Ferreira, M. G. S. Zn–Al layered double hydroxides as chloride nanotraps in active protective coatings. *Corros. Sci.*, **2012**, *55,* 1–4, DOI: 10.1016/j.corsci.2011.10.003.

27. Maia, F.; Tedim, J.; Lisenkov, A. D.; Salak, A. N.; Zheludkevich, M. L.; Ferreira, M. G. S. Silica nanocontainers for active corrosion protection. *Nanoscale*, **2012**, *4*, 1287-1298, DOI: 10.1039/C2NR11536K.

28. Eiby, S. H. J.; Tobler, D. J.; Nedel, S.; Bischoff, A.; Christiansen, B. C.; Hansen, A. S.; Kjaergaard, H. G.; Stipp, S. L. S. Competition between chloride and sulphate during the reformation of calcined hydrotalcite. *Appl. Clay Sci*. **2016**, *132–133*, 650–659, DOI:10.1016/j.clay.2016.08.017.

29. Miyata, S. Anion-exchange properties of hydrotalcite-like compounds. *Clays and Clay Miner.*, **1983**, *31*, 305-311, DOI: 10.1346/CCMN.1983.0310409.

30. Evans, D. G.; Slade, R. C. T. Structural Aspects of Layered Double Hydroxides. In *Layered double hydroxides;* Duan, X., Evans, D., Eds.; Springer-Verlag Berlin, Heidelberg, 2006, 119, pp. 1–87, DOI: 10.1007/430_005.

31. Mikhailau, A.; Maltanava, H. M.; Poznyak, S. K.; Salak, A. N.; Zheludkevich, M. L.l; Yasakau, K. A.; Ferreira, M. G. S. One-step synthesis and growth mechanism of nitrate intercalated ZnAl LDH conversion coatings on zinc. *ChemComm.* **2019**, *55,* 6878-6881, DOI: 10.1039/C9CC02571E.

32. Seeck, O. H.; Deiter, C.; Pflaum, K.; Bertam, F.; Beerlink, A.; Franz, H.; Horbach, J.; Schulte-Schrepping, H.; Murphy, B. M.; Greve, M.; Magnussen, O. The high-resolution diffraction beamline P08 at PETRA III, *J. Synchrotron.Rad*. 19 (2012) 30-38, DOI: 10.1107/S0909049511047236.

33. Bouali, A. C.; Iuzviuk, M. H.; Serdechnova, M.; Yasakau, K. A.; Wieland, D. C. F.; Dovzhenko, G. Ferreira, M. G. S.; Zobkalo, I. A.; Zheludkevich, M. L. Crystal structure and kinetics comparison of Zn-Al LDH grown on AA2024-T3 and pure zinc and their intercalation with chloride. *Applied Surface Science,* **2020**, *501,* 144027, DOI: 10.1016/j.apsusc.2019.144027.

34. Casas-Cabanas, M.; Reynaud, M.; Rikarte, J.; Horbach, P.; Rodríguez-Carvajal, J. FAULTS: a program for refinement of structures with extended defects, *J. Appl. Crystallogr.*, **2016**, *49*, 2259-2269, DOI: 10.1107/S1600576716014473.

35. Rowe, M. C.; Brewer, B. J. AMORPH: A statistical program for characterizing amorphous materials by X-ray diffraction, *Comput & Geosci.,* **2018**, *120*, 21-31, DOI: 10.1016/j.cageo.2018.07.004.

36. Marappa, S.; Radha, S.; Kamath, P. V. Nitrate-Intercalated Layered Double Hydroxides–Structure Model, Order, and Disorder. *Eur. J. Inorg. Chem.,* **2013**, *2013*, 2122-2128, DOI: 10.1002/ejic.201201405.

37. Williams, G. R.; Khan, A. I.; O'Hare, D. Mechanistic and Kinetic Studies of Guest Ion Intercalation into Layered Double Hydroxides Using Time-Resolved, In situ Powder Diffraction. In *Layered double hydroxides;* Duan, X., Evans, D., Eds.; Springer-Verlag Berlin, Heidelberg, 2006, 119, pp. 161–192, DOI: 10.1007/430_002.

38. Serdechnova, M.; Salak, A. N.; Barbosa, F. S.; Vieira, D. E. L.; Tedim, J.; Zheludkevich, M. L.; Ferreira, M. G. S. Interlayer intercalation and arrangement of 2-mercaptobenzothiazolate and 1,2,3-benzotriazolate anions in layered double hydroxides: In situ X-ray diffraction study, *J. Solid State Chem.*, **233** (2016) 158–165, DOI: 10.1016/j.jssc.2015.10.023.

39. Avrami, M. Kinetics of Phase Change. I General Theory. *J. Chem. Phys.*, **1939**, *7*, 1103–1112, DOI: 10.1063/1.1750380.

40. Avrami, M. Kinetics of Phase Change. II Transformation-Time Relations for Random Distribution of Nuclei. *J. Chem. Phys.*, **1940**, *8*, 212–224, DOI: 10.1063/1.1750631.

41. Avrami, M. Granulation, Phase Change, and Microstructure Kinetics of Phase Change. III. *J. Chem. Phys.*, **1941**, *9*, 177–184, DOI: 10.1063/1.1750872.

42. Erofeev, B. V. A generalized equation of chemical kinetics and its application in reactions involving solids, *Comptes Rendus de L'Academie des Sciences de l'URSS*, **1946**, *52*, 511-514.

43. Hancock, J. D.; Sharp, J. H. Method of comparing solid-state kinetic data and its application to the decomposition of kaolinite, brucite, and $BaCO_3$. *J. Am. Ceram. Soc.*, **1972** *55*, 74-77, DOI: 10.1155/2014/710487.

44. Hulbert, S. F. "Models for solid-state reactions in powdered compacts: A review. *J. Br. Ceram. Soc.*, **1969**, *6*, 11-20.

45. Twu, J.; Dutta, P. K. Structure and Reactivity of Oxovanadate Anions in Layered Lithium Alumínate Materials. *J. Phys. Chem.*, **1989**, *93,* 7863-7868, DOI: 10.1021/j100360a028.

46. Auerbach, S. M.; Carrado, K. A.; Dutta, P. K. Handbook of layered materials**.** Taylor & Francis, 2004, pp. 1-664, DOI: 10.1201/9780203021354.

47. Barriga, C.; Jones, W.; Malet, P.; Rives, V.; Ulibarri, M. A. Synthesis and Characterization of Polyoxovanadate-Pillared Zn−Al Layered Double Hydroxides: An X-ray Absorption and Diffraction Study. *Inorg. Chem.*, **1998,** *37*, 1812-1820, DOI: 10.1021/ic9709133.

48. Brindley, G. W.; Kao, C.-C. Structural and IR Relations Among Brucite-Like Divalent Metal Hydroxides. *Phys. Chem. Minerals*, **1984**, *10*, 187-191, DOI: 10.1007/BF00311476.